# Imaging and characterization of conducting ferroelectric domain walls by photoemission electron microscopy

J. Schaab[1], I. P. Krug[2,3], F. Nickel[3], D. M. Gottlob[3], H. Doğanay[3], A. Cano[4], M. Hentschel[5,6], Z. Yan[6], E. Bourret[6], C. M. Schneider[3], R. Ramesh[6,7], and D. Meier[1,*]

[1]Department of Materials, ETH Zürich, Vladimir-Prelog-Weg 4, 8093 Zürich, Switzerland

[2]Institut für Optik und Atomare Physik, TU Berlin, Hardenbergstrasse 36, 10623 Berlin, Germany

[3]Peter Grünberg Institute (PGI-6), Leo-Brandt-Strasse, 52425 Jülich, Germany

[4]CNRS, University Bordeaux, ICMCB, UPR 9048, F-33600 Pessac, France

[5]4th Physics Institute and Research Center SCoPE, University of Suttgart, Pfaffenwaldring 57, 70659 Stuttgart, Germany

[6]Materials Sciences Division, Lawrence Berkeley National Laboratory, Berkeley, California 94720, USA

[7]Department of Materials Science and Engineering, University of California, Berkeley, California 94720, USA

*Corresponding author: dennis.meier@mat.ethz.ch

Abstract: High-resolution X-ray photoemission electron microscopy (X-PEEM) is a well-established method for imaging ferroelectric domain structures. Here, we expand the scope of application of X-PEEM and demonstrate its capability for imaging and investigating domain walls in ferroelectrics with high-spatial resolution. Using ErMnO$_3$ as test system, we show that ferroelectric domain walls can be visualized based on photo-induced charging effects and local variations in their electronic conductance can be mapped by analyzing the energy distribution of photoelectrons. Our results open the door for non-destructive, contract-free, and element-specific studies of the electronic and chemical structure at domain walls in ferroelectrics.

Domain walls in ferroelectric and multiferroic materials currently attract broad attention due to the anomalous and functional physical properties that emerge at this natural type of oxide interface [1-3]. During the last decade such domain walls were demonstrated to promote photovoltaic effects [4], enhanced electromechanical response [5], anomalous electronic transport behavior [6-13], magnetoresistive properties [14], and more (see e.g. [15] for a review). Despite all the activity in the research field and the remarkable progress that has been made in experiment and theory [16-19], the experimental characterization of intrinsic domain-wall properties remains a challenging task. Scanning probe microscopy (SPM) has evolved the most common technique for gaining insight to the physics of ferroelectric domain walls and related effects. Today, SPM variants like conductive atomic force microscopy (c-AFM) and piezoresponse force microscopy (PFM) are routinely used to measure the local electronic transport [15] and piezoresponse [20] at these domain walls. Besides SPM, mainly electron microscopy techniques such as transmission electron microscopy (TEM) or scanning electron microscopy (SEM) have been applied to gain valuable information about the structure [21-24] and e.g. dielectric permittivity [25-27] at ferroelectric and multiferroic domain walls. We are, however, only at the verge of understanding the complex nano-physics of these functional oxide interfaces and a further expansion of the accessible parameter space is highly desirable.

In this work we apply high-resolution X-ray photoemission electron microscopy (X-PEEM) to ferroelectric domain walls and demonstrate the general feasibility of X-PEEM for domain-wall studies, as well as new opportunities it offers for characterizing such functional nano-objects. In our ferroelectric test system, ErMnO$_3$, we visualize domain walls by X-PEEM and map their anomalous electronic conductance properties contact-free by analyzing the energy distribution of photo-excited electrons. Because X-PEEM has not yet been used for studying ferroelectric domain walls – although it is well-established for imaging domains in ferroelectrics [28-30] or ferroelectric domains in multiferroics [31,32] – we also compare our measurements to conventional c-AFM and PFM scans. Our results show that X-PEEM is sensitive to the anomalous transport behavior emerging at ferroelectric domain walls and reveal a new pathway for non-destructive and element-specific studies of their chemical and electronic structures.

For our X-PEEM domain-wall experiments we used ErMnO$_3$ as test system because it is ferroelectric at room-temperature [33], exhibits all proto-types of ferroelectric domain walls (neutral, head-to-head, and tail-to-tail), and is well characterized in terms of SPM and theory [9,11,34,35]. High-quality ErMnO$_3$ single-crystals were grown by the floating-zone method, oriented by Laue diffraction, and cut into platelets with the spontaneous ferroelectric polarization ($P \parallel z$) lying in the surface plane (x-cut). The prepared samples had lateral dimensions of about 4 mm and a thickness of 1 mm. To achieve the flat surfaces required for our measurements, we chemo-mechanically polished the ErMnO$_3$ platelets using silica slurry.

In Fig. 1(a) we present the ferroelectric domain structure of our ErMnO$_3$ sample which was imaged measuring the in-plane PFM response under ambient conditions. The PFM image reveals the typical $R$MnO$_3$ ($R$ = Sc, Y, In, Dy–Lu) domain pattern with its characteristic six-fold meeting points composed of alternating +P and −P domains [36,37]. To develop a coordinate system that allows for reproducibly finding certain surface positions, Pt-markers with a size of 100 x 100 µm² were designed using electron-beam lithography. The markers are visible in the upper and lower left corner of the PFM scan, as well as in the inset to Fig. 1(a) which shows an enlarged PEEM image gained by illuminating the sample with an Hg-lamp (Hg-PEEM).

Figure 1(b) displays a zoom-in to the ferroelectric domain structure in Fig. 1(a) (in-plane PFM). The corresponding c-AFM data is presented in Fig. 1(c). This data set evidences that the $ErMnO_3$ crystal exhibits the same electronic domain-wall properties as previously reported [9], i.e. insulating head-to-head and conducting tail-to-tail domain walls at a DC bias of -4 V applied to the tip. An X-PEEM image taken at the same sample position and photon energy of 641.5 eV (Mn $L_3$ edge) is shown in Fig. 1(d). Here, two dark lines clearly distinguish from an otherwise homogeneously grey background. A comparison with the PFM and c-AFM scans in Figs. 1(b) and (c) identifies these dark lines as electrically conducting tail-to-tail domain walls. In the X-PEEM data, however, only the tail-to-tail walls are visible while insulating head-to-head domain walls are indistinguishable from domain areas.

To better understand the mechanism responsible for the obtained X-PEEM contrast, we performed additional measurements as summarized in Fig. 2. Figure 2(a) shows an X-PEEM image of a tail-to-tail wall (641.5 eV) taken with optimized contrast. Here, maximum brightness of domain areas was achieved by fine-tuning the voltage $U_a = U_0 +$ $U_{st}$, which accelerates emitted electrons, so that primarily those from bulk regions can pass the microscope`s energy filter ($U_0$ = 15 keV, −10 eV ≲ $U_{st}$ ≲ +10 eV). We further found that the contrast was inverted when reducing the acceleration voltage as a comparison of Figs. 2(a) and 2(b) reveals. This behavior indicates that photo-excited electrons from domain-wall regions have a markedly higher kinetic energy, $E_{kin}$, than those from the bulk. In consequence, they are able to pass the energy filter even at reduced acceleration voltage, whereas photoelectrons from the bulk are largely blocked leading to the bright walls in Fig. 2(b). The difference in $E_{kin}$ can directly be seen in the dispersive plane of the microscope as shown in Fig. 2(c) [38].

Figure 2(c) provides insight to the photoelectron distribution in k-space when illuminating the sample with X-rays at an energy of 641.5 eV. The two-dimensional plot displays a projection of the photoelectron distribution onto the dispersive $E_{kin}$-$k_y$ plane, $I(E_{kin},k_y)$, for a single $k_x$ with the latter being selected by the energy filter settings as explained in ref. [38]. A typical distribution is limited by a parabola due to the $\sqrt{E_{kin}}$-dependence of the size of the Ewald sphere as sketched in the inset to Fig. 2(c). In the present case, however, two parabolas are clearly distinguishable with the low-energy parabola being associated to electrons emitted from the bulk. The corresponding electron energy distribution is sketched in Fig. 2(d) where we illustrate the electron yield as function of energy [39]. Depending on the energy-filter settings (indicated by the two colored bars) either bulk or domain-wall photoelectrons can pass which explains the energy contrasts in Figs. 2(a) and 2(b), respectively.

In order to quantify the observed difference in kinetic energy $\Delta E$ (see Fig. 2(d)) we measured the electron yield as function of the so-called start energy $U_{st}$ and analyzed local intensity variations. The result gained for the domain-wall cross-section marked in Fig. 2(b) is presented in Fig. 3(a) where we plot the secondary electron yield against position and $U_{st}$. The three-dimensional plot reveals that the electron-energy distribution at the domain wall is shifted towards higher energies with $\Delta E \approx 1$ eV. This shift is also evident in the associated projection onto the xy-plane showing lines of equal intensity.

A complete photoelectron-energy map for the domain walls in Figs. 2(a) and 2(b) is depicted in Fig. 3(b). The map presents the peak position of the energy distribution (averaged over 2 x 2 pixels) with higher energy values corresponding to faster electrons. In addition to the above discussed difference between domain wall and bulk regions, the spatially resolved data uncovers a correlation between $\Delta E$ and the domain-wall orientation. The latter is reminiscent of the orientation-dependent electronic conductance observed by cAFM [9] and suggests a connection between the domain-wall transport properties and the emergent energy contrasts in X-PEEM, which we will discuss in the following.

The illumination with intense X-rays leads to photo-excitation of charge carriers (electrons). In case of an insulating or poorly conducting ferroelectric, like the $ErMnO_3$ test system, the bulk material cannot compensate for the emitted photoelectrons and hence gets positively charged. Due to the positive charging photoelectrons get slowed down and the associated energy distribution shifts to lower energies. At the tail-to-tail walls, however, charging effects are largely suppressed because of the locally enhanced conduction properties that allow compensating for the photo-induced charging. This difference in electronic transport can explain the presence of two maxima in the energy distribution and resulting contrasts, as well as the angular dependence evident in Fig. 3(b). Thus, we conclude that photo-induced charging is responsible for the X-PEEM contrast at the conducting tail-to-tail domain walls. We note that domain wall contrasts only occur at the Mn $L_3$ edge where the absorption coefficient is high and a large number of photoelectrons are emitted. No difference between bulk and tail-to-tail domain walls is observed at lower energies, i.e. before the Mn $L_3$ edge. This observation excludes possible work-function differences as the source of the domain-wall contrast in X-PEEM.

To further support the conclusion that the observed X-PEEM contrasts emerge due to photo-induced charging effects, we performed imaging experiments with variable synchrotron beam intensity. For this experiment the beam intensity and profile width was controlled by changing the width of the entrance slit that X-rays pass before reaching the experiment. The result is presented in Fig. 3(c) and shows a striking dependence of the domain-wall contrast, $\frac{I_{wall}}{I_{bulk}}$, on the synchrotron intensity which is in tune with the above interpretation of the X-PEEM contrasts: At low intensity (region 1), the electron yield in bulk areas is higher compared to the tail-to-tail walls because mobile holes accumulate at this type of domain wall and hence electrons that may be emitted are rare. At higher intensity (region 2), however, the sample drastically charges positively so that local differences in conductance dominate the X-PEEM contrast leading to a crossover from $\frac{I_{wall}}{I_{bulk}} < 1$ to $\frac{I_{wall}}{I_{bulk}} > 1$. The intensity-dependent measurement highlights that domain-wall contrasts can be improved using a higher X-ray intensity. Vice versa, unwanted charging effects can be largely suppressed by limiting the beam intensity for e.g. recording reliable domain-wall spectra after detecting its position.

In summary, we showed that conducting domain walls in a ferroelectric bulk material can be visualized using high-resolution X-PEEM and we explained the emergent contrasts based on photo-induced charging effects caused by intense X-ray illumination. The charging contrast, in its turn, is basically determined by the local conductivity state at the domain wall. The X-PEEM method offers data acquisition times in the order of 0.1 – 10 s which is significantly faster compared to conventional SPM scans. Moreover, domain-wall related anomalies and subtle variations in electronic conductance can be detected contact-free and hence without contributions from contact resistance by analyzing the energy distribution of photo-excited electrons. The demonstrated ability of X-PEEM to directly access domain walls opens the door for element-specific investigations that can provide access to so far unexplored ferroelectric domain-wall properties such as chemical structure, local valence states, and symmetry violations.

Acknowledgement: The authors thank M. Fiebig, Y. Kumagai, and N. A. Spaldin for fruitful discussions, M. Lilienblum for technical assistance, and the BESSY staff and A.

Kaiser (Specs GmbH) for support. D.M. acknowledges financial support by the Alexander von Humboldt Foundation.

References

1. E. K. H. Salje, ChemPhysChem **11**, 940 (2010).

2. E. K. H. Salje and H. Zhang, Phase Transit. **82**, 452 (2009).

3. H. Bea and P. Paruch, Nat. Mater. **8**, 168 (2009).

4. Y. S. Yang, J. Seidel, S. J. Byrnes, P. Shafer, C.-H. Yang, M. D. Rossell, P. Yu, Y.-H. Chu, J. F. Scott, J. W. Ager III, L. W. Martin, and R. Ramesh, Nat. Nanotechnol. **5**, 143 (2010).

5. T. Sluka, A. K. Tagantsev, D. Damjanovic, M. Gureev, and N. Setter, Nat. Commun. **3**, 748 (2012).

6. J. Seidel, L. W. Martin, Q. He, Q. Zhan, Y.-H. Chu, A. Rother, M. E. Hawkridge, P. Maksymowych, P. Yu, M. Gajek, N. Balke, S. V. Kalinin, S. Gemming, F. Wang, G. Catalan, J. F. Scott, N. A. Spaldin, J. Orenstein, and R. Ramesh, Nat. Mater. **8**, 229 (2009).

7. J. Guyonnet, I. Gaponenko, S. Gariglio, and P. Paruch, Adv. Mater. **23**, 5377 (2011).

8. S. Farokhipoor and B. Noheda, Phys. Rev. Lett. **107**, 127601 (2011).

9. D. Meier, J. Seidel, A. Cano, K. Delaney, Y. Kumagai, M. Mostovoy, N. A. Spaldin, R. Ramesh, and M. Fiebig, Nat. Mater. **11**, 284 (2012).

10. P. Maksymovych, A. N. Morozovska, P. Yu, E. A. Eliseev, Y.-H. Chu, R. Ramesh, A. P. Baddorf, and S. V. Kalinin, Nano Lett. **12**, 209 (2012).

11. W. Wu, Y. Horibe, N. Lee, S.-W. Cheong, and J. R. Guest, Phys. Rev. Lett. **108**, 077203 (2012).

12. M. Schröder, A. Haußmann, A. Thiessen, E. Soergel, T. Woike, and L. M. Eng, Adv. Funct. Mater. **22**, 3936 (2012).

13. T. Sluka, A. K. Tagantsev, P. Bednyakov, and N. Setter, Nat. Commun. **4**, 1808 (2013).

14. Q. He, C.-H. Yeh, J.-C. Yang, G. Singh-Bhalla, C.-W. Liang, P.-W. Chiu, G. Catalan, L. W. Martin, Y.-H. Chu, J. F. Scott, and R. Ramesh, Phys. Rev. Lett. **108**, 067203 (2012).

15. G. Catalan, J. Seidel, R. Ramesh, and J. F. Scott, Rev. Mod. Phys. **84**, 119 (2012).

16. B. Meyer and D. Vanderbilt, Phys. Rev. B **65**, 104111 (2002).

17. J. Padilla, W. Zhong, and D. Vanderbilt, Phys. Rev. B **53**, R5969 (1996).

18. E. A. Eliseev, A. N. Morozovska, G. S. Svechnikov, V. Gopalan, and V. Y. Shur, Phys. Rev. B **83**, 235313 (2011).

19. M. Y. Gureev, A. K. Tagantsev, and N. Setter, Phys. Rev. B **83**, 184104 (2011).

20. E. B. Lochocki, S. Park, N. Lee, S.-W. Cheongm and W. Wu, Appl. Phys. Lett. 99, 232901 (2011).

21. C.-L. Jia, S.-B. Mi, K. Urban, I. Vrejoiu, M. Alexe, and D. Hesse, Nat. Mater. **7**, 57 (2008).

22. M.-G. Han, Y. Zhu, L. Wu, T. Aoki, V. Volkov, X. Wang, S. C. Chae, Y. S. Oh, and S.-W. Cheong, Adv. Mater. **25**, 2415 (2013).

23. Q. Zhang, G. Tan, L. Gu, Y. Yao, C. Jin, Y. Wang, X. Duan, and R. Yu, Sci. Rep. **3**, 2741 (2013).

24. T. Matsumoto, R. Ishikawa, T. Tohei, H. Kimura, Q. Yao, H. Zhao, X. Wang, D. Chen, Z. Cheng, N. Shibata, and Y. Ikuhara, Nano Lett. **13**, 4594 (2013).

25. V. V. Aristov, L. S. Kokhanchik, K.-P. Meyer, and H. Blumtritt, Phys. Status Solidi A **78**, 229 (1983).

26. V. V. Aristov, L. S. Kokhanchik, and Y. I. Voronovskii, Phys. Status Solidi A **86**, 133 (1984).

27. J. Li, H. X. Yang, H. F. Tian, C. Ma, S. Zhang, Y. G. Zhao, and J. Q. Li, Appl. Phys. Lett. **100**, 152903 (2012).

28. E. Arenholz, G. van der Laan, A. Fraile-Rodríguez, P. Yu, Q. He, and R. Ramesh, Phys. Rev. B **82**, 140103 (2010).

29. I. Krug, N. Barrett, A. Petraru, A. Locatelli, T. O. Mentes, M. A. Niño, K. Rahmanizadeh, G. Bihlmayer, and C. M. Schneider, Appl. Phys. Lett. **97**, 222903 (2010).

30. N. Barrett, J. E. Rault, J. L. Wang, C. Mathieu, A. Locatelli, T. O. Mentes, M. A. Niño, S. Fusil, M. Bibes, A. Barthélémy, D. Sando, W. Ren, S. Prosandeev, L. Bellaiche, B. Vilquin, A. Petraru, I. P. Krug, and C. M. Schneider, J. Appl. Phys. **113**, 187217 (2013).

31. S. Cherifi, R. Hertel, S. Fusil, H. Béa, K. Bouzehouane, J. Allibe, M. Bibes, and A. Barthélémy, Phys. Status Solidi RRL **4**, 22 (2010).

32. J. E. Rault, W. Ren, S. Prosandeev, S. Lisenkov, D. Sando, S. Fusil, M. Bibes, A. Barthélémy, L. Bellaiche, and N. Barrett, Phys. Rev. Lett. **109**, 267601 (2012).

33. G. A. Smolenskiĭ and I. E. Chupis, Soviet Phys. Usp. **25**, 475 (1982).

34. D. Meier, M. Lilienblum, P. Becker, L. Bohatý, N. A. Spaldin, R. Ramesh, and M. Fiebig, Phase Transit. **86**, 33 (2013).

35. Y. Kumagai and N. A. Spaldin, Nat. Commun. **4**, 1540 (2013).

36. T. Choi, Y. Horibe, H. T. Yi, Y. J. Choi, W. Wu, and S.-W. Cheong, Nat. Mater. **9**, 253 (2010).

37. T. Jungk, Á. Hoffmann, M. Fiebig, and E. Soergel, Appl. Phys. Lett. **97**, 012904 (2010).

38. R. M. Tromp, Y. Fujikawa, J. B. Hannon, A. W. Ellis, A. Berghaus, and O. Schaff, J. Phys. : Condens. Matter **21**, 314007 (2009).

39. P. W. Hawkes and J. C. Spence (eds.), *Science of Microscopy.* Springer Science + Business Media, LLC (2007).

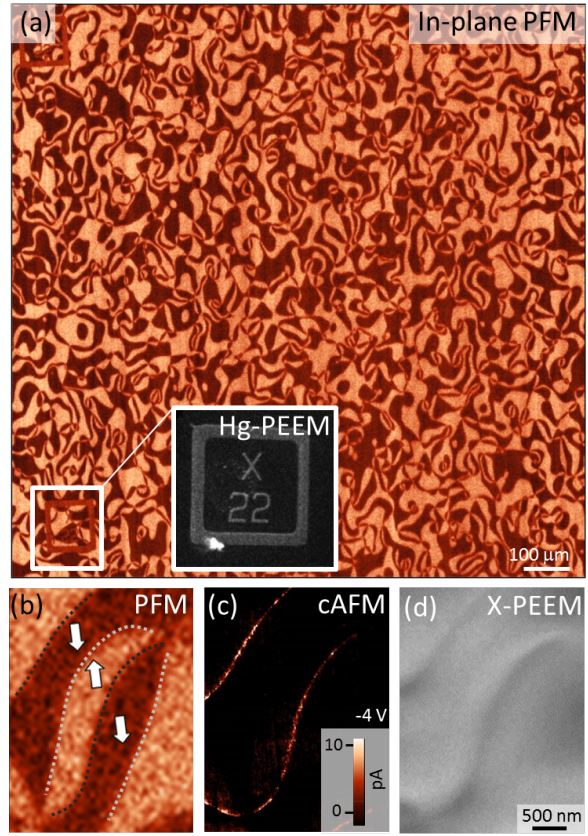

**Figure 1:** (a) PFM image (in-plane contrast) of the ferroelectric domain structure in ErMnO$_3$ with the spontaneous polarization lying in the plane (x-cut). Pt-markers on the surface, as shown by the Hg-PEEM image in the inset to (a), allowed for investigating the same sample position by different microscopy methods. Figs. (b), (c), and (d) compare PFM, c-AFM, and X-PEEM data as detailed in the main text.

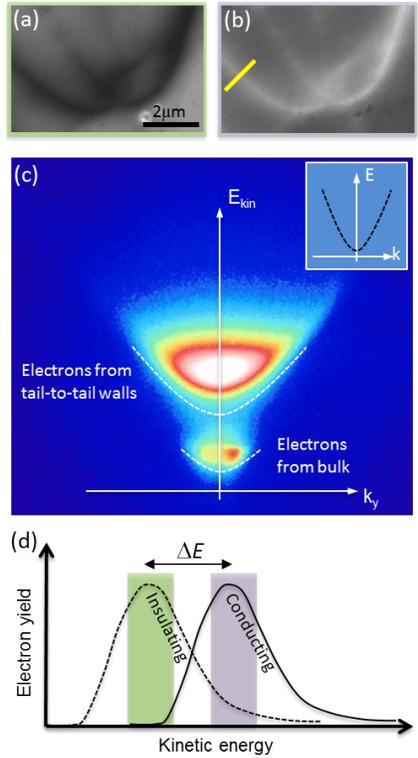

**Figure 2:** (a), (b) X-PEEM images of tail-to-tail domain walls taken on ErMnO$_3$ at a photon energy of 641.5 eV (Mn L$_3$ edge). Contrasts are inverted for the two images due to different analyzer settings of the microscope as sketched in (d). (c) Kinetic energy distribution of emitted photoelectrons when illuminating the sample with X-rays at 641.5 eV as seen in the dispersive plane of the microscope. High and low intensity levels are color-coded white and blue, respectively, and white dotted lines highlight the presence of two distinct parabolas as detailed in the text. (d) Schematic illustration of the energy distribution of photoelectrons emitted from insulating and conducting sample areas. Color-coded boxes indicate energy filter settings that yield contrasts as observed in (a) and (b).

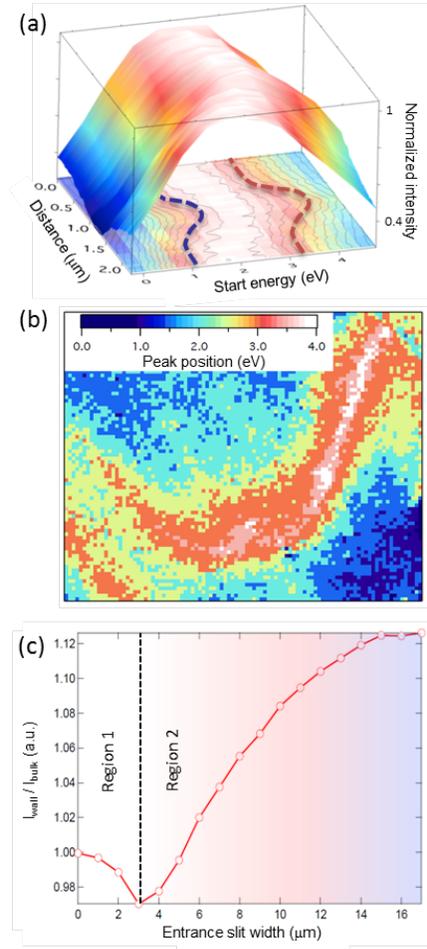

**Figure 3:** (a) Evolution of the photoelectron distribution along the domain-wall cross-section marked in 2(b). At the position of the domain wall the energy distribution is shifted towards higher energies by about 1 eV. (b) Conductivity map derived from the X-PEEM data shown in 2(a). The color code reflects the peak position of the photoelectron distribution in eV as shown in the inset to (b). (c) Dependence between domain-wall contrasts and X-ray intensity.